\documentclass[sigconf, authorversion, nonacm]{acmart}

\usepackage[T1]{fontenc}
\usepackage{booktabs}
\usepackage{threeparttable}
\usepackage{caption}

\renewcommand\footnotetextcopyrightpermission[1]{} 
\usepackage{xurl}
\usepackage{booktabs}
\usepackage{subcaption}
\usepackage{tabularx}
\usepackage{enumitem}

\author{Rebecca Umbach}
\authornote{Authors contributed equally to this research.}
\affiliation{\institution{Google} \city{Mountain View} \country{USA}}
\email{rumbach@google.com}

\author{Griffin Hunt}
\authornotemark[1]
\affiliation{\institution{Google} \city{Mountain View} \country{USA}}

\author{John Buckley}
\affiliation{\institution{The LEGO Group} \city{Billund} \country{Denmark}}

\author{Joel Scanlan}
\affiliation{\institution{CSAM Deterrence Centre, University of Tasmania} \city{Hobart} \country{Australia}}

\author{Caoilte Ó Ciardha}
\affiliation{\institution{University of Kent} \city{Canterbury} \country{UK}}

\author{Ethel Quayle}
\affiliation{\institution{University of Edinburgh} \city{Edinburgh} \country{UK}}

\author{Ainslie Heasman}
\affiliation{\institution{Centre for Addiction and Mental Health} \city{Toronto} \country{Canada}}

\author{Maximilian von Heyden}
\affiliation{\institution{Charité - Universitätsmedizin Berlin
} \city{Berlin} \country{Germany}}

\author{Elizabeth Letourneau}
\affiliation{\institution{Johns Hopkins University} \city{Baltimore} \country{USA}}

\author{Donald Findlater}
\affiliation{\institution{Lucy Faithfull Foundation} \city{Epsom} \country{UK}}
\affiliation{\institution{Stop It Now! UK and Ireland} \city{Epsom} \country{UK}}

\author{Tegan Insoll}
\affiliation{\institution{Protect Children} \city{Helsinki} \country{Finland}}

\author{Richard Wortley}
\affiliation{\institution{University College London} \city{London} \country{UK}}

\author{Chad Steel}
\affiliation{\institution{George Mason University} \city{George Mason} \country{USA}}

\author{Abhishek Roy}
\authornotemark[1]
\affiliation{\institution{Google} \city{Mountain View} \country{USA}}
\renewcommand{\shortauthors}{Umbach et al.}
\title{Deterring Searches for Child Sexual Abuse Material on Google Search and Promoting Help-Seeking}

 \begin{abstract}
 Google Search deploys a ``Onebox'' feature at the top of the results page when users conduct searches for Child Sexual Abuse Material. This study evaluates the impact of a strategic shift in this feature, comparing a revised intervention, focused on repercussions and therapeutic resources, to a previous iteration that focused on reporting. Using a difference-in-differences analysis of internal Google Search logs data, we found the new messaging resulted in a 3.8 percentage point reduction as compared to the status quo in subsequent CSAM-related queries within the same Search session. We found an average click through rate of 0.73\% on any of the hyperlinked buttons to help-providing resources. Together, this research presents convergent evidence that a subset of individuals can be deterred from ongoing CSAM-seeking and redirected to therapeutic services.
 \end{abstract}

\begin{document}

\maketitle



\section{Introduction}
\label{introduction}

Growing global access to the internet has facilitated the production, distribution, and consumption of child sexual abuse material (CSAM)\citep{smallbone2017preventing, napier2024characteristics}. 
CSAM causes significant and long-term harms to victims \citep{donevan2025experience, finkelhor2024prevalence, chauvire2024victims, page2025psychological}, as well as those close to individuals who have offended or are at risk of offending \citep{armitage2024we, kavanagh2024your}.
In 2024, the CyberTipline\footnote{The reporting tipline in the United States run by the National Center for Missing and Exploited Children. Despite being an American-based organization, it acts as a clearinghouse and liaises with law enforcement globally.} received over 20 million reports, documenting activity in practically every country\footnote{Any estimate of the scale of this problem  is likely to suffer bi-directional biases. Electronic communication service providers and remote computing service providers must report apparent CSAM to the CyberTipline by U.S. federal law. In effect, this leads to duplicative reports and false positive reports, the negative effects of which are well documented \citep{stanford_cybertipline_2024}. This figure also overlooks abusive material intentionally hidden using encryption or the anonymity of the dark web.}  \citep{ncmec2025} and research has identified a rise in reports of CSAM \citep{bursztein2019rethinking, gannoni2023preventing, alliance2019global}.  
Electronic service providers (ESPs) including social media and messaging platforms \citep{teunissen2022child}, legal pornography websites \citep{morgan2018understanding}, and websites located by search engines \citep{steel2015web, westlake2017assessing, steel2022technical} remain central to how CSAM is accessed and shared online. 
It has been found that the majority of self-reported CSAM offenders report first encountering the illegal CSAM by accident on the open web \citep{insoll2024tech, napier2024characteristics, wortley2024accessing}, highlighting the vital role of mainstream online services to ensure their platforms are active in removing the content and disrupting attempts to access it.

ESPs can and do take significant steps to prevent the spread and presence of CSAM online. These include hashing CSAM imagery to facilitate automated detection and removal \citep{google:hashmatching, farid2021overview,  Meta2025OnlineChild}, directing to reporting options such as the National Center for Missing and Exploited Children (NCMEC) and the Internet Watch Foundation (IWF), and collaborative efforts with other tech companies to share signals for more effective enforcement [e.g., Project Lantern \citep{techcoalition:lantern}].  Google Search also builds and deploys classifiers and ranking protections to eliminate risky content for CSAM-seeking queries and promote high quality results, such as educational or news-related links \citep{google_csam_2025}.
These interventions help address the supply side of CSAM, but there is an urgent need to scale  
prevention efforts to better help address the demand side. 
ESPs are uniquely well-positioned to respond to users seeking CSAM at the exact time of the problematic behavior - potentially at the time of an involvement decision, where it is likely to be most effective \citep{clarke2017routine}. They have an opportunity to deliver just-in-time, targeted information to deter the immediate behavior and divert the individual to effective therapeutic services \citep{insoll2024factors, shields2020help, cahill2025self}. This study evaluates whether an ESP-delivered intervention successfully 1) deterred individuals from seeking CSAM and 2) redirected individuals to robust therapeutic resources.

In answering these questions, the current study uniquely examines large-scale, multi-country data comprising Google user log data and helpline web traffic data. 

\section{Background}
\label{background}
 While the creation and distribution of CSAM predates the internet, the internet and other novel technologies such as artificial intelligence have transformed the volume \citep{bursztein2019rethinking} and content \citep{Salter02092022, pfefferkorn2025addressing}, and have reduced barriers to accessing CSAM \citep{quayle2004child, wortley2006child, price2024review}. Digital environments facilitate child exploitation and the anonymous exchange of CSAM \citep{wang2023investigating, huikuri2023users, lahtinen2025investigating}, but CSAM also proliferates on the open web through, for example, poor content moderation practices \citep{mckee2022pornhub}.
Research has increasingly worked to frame child sexual abuse as a preventable public health problem \citep{letourneau2014need,fix2025messaging}, offering a framework that accommodates interventions spanning three levels of prevention \citep{brantingham1976conceptual, price2024review}, and which can be extended to online offending (i.e., CSAM). Broadly, this framework includes efforts to  target whole populations to prevent the onset of harm, interventions aimed at at-risk individuals to prevent the onset of a behaviour, and interventions to prevent re-offense and support survivors of child sexual abuse. 

Two areas of prevention research are of particular relevance to this study. A subset of prevention research has focused on digital interventions designed to deter would-be online offenders  through increasing perceived risks and making CSAM discovery harder \citep{price2024review, scanlan2022creating}. The underpinning theory behind these types of efforts is Routine Activities Theory, which posits that crimes occur when a motivated offender encounters a suitable target in the absence of a capable guardian \citep{cohen1979social}. By increasing surveillance, or the perception of surveillance, ESP interventions can act as a digital guardian, reducing the likelihood of offending. 

A complementary body of research has focused on intervening at the individual level, including the development of effective therapeutic interventions and barriers for help-seeking, including both individual factors (e.g., shame, fear of consequences, self-hatred) and societal factors (e.g., stigma, lack of access to competent help) \citep{levenson2017obstacles, harper2018reducing, kothari2021understanding, chronos2024treatment, wortley2024accessing, cahill2025self, wolbers2024drivers}. 

\subsection{Interventions}
\label{interventionsection}
Theories around deterrence of CSAM-seeking come primarily from observational studies (e.g., analyzing forum conversations between individuals with self-reported sexual interest in children) \citep{murphy2024pilot}, self-report studies (e.g., surveys or interviews with same) \citep{ociardha2025iatso}, or honeypot (field) experimental studies \citep{prichard2024effect}. Across these different studies, there is general agreement that interventions can help reduce the likelihood of seeking out CSAM \citep{wortley2024accessing}. 
\subsubsection{Clinician/staff-led, self-directed}
\label{clinician}

Increasingly, civil society organizations are establishing dedicated CSAM prevention programs, which may incorporate anonymous chat or helpline capabilities, therapy and/or self-help materials \citep{shields2020help, grant2019didn, schuler2021characteristics, seto2024evaluating, newman2024impact, beier2015german}. Programs focus on assisting youth and/or adults who are distressed about their sexual attraction to children and/or their online or offline behaviour involving children. Some governments fund child sexual abuse and CSAM prevention programs as part of their health or justice (i.e., crime prevention) portfolio [e.g., Talking for Change in Canada \citep{talkingforchange} and Kein Täter werden in Germany \citep{keintaeterwerden}], which allows for more sustainable and embedded national investment in child protection. Typically designed and run by clinicians and experts, these programs often undergo evaluation and have appropriate protections in place to safeguard those seeking help, as well as the service providers \citep{shields2020help, schuler2021characteristics, latth2022effects}. Web-based initiatives reduce barriers for seeking help, and evaluations show promising results both in reducing recidivism and preventing initial offenses \citep{hillert2024web}, although additional rigorous evaluations are needed.

\subsubsection{Peer-to-peer interventions}
There are online support groups or networks (which may be moderated or clinically supported) where peers, who identify with a sexual attraction to children, support each other in nonoffending \citep{cantor2016non, nielsen2022virtuous, jones2021identifying, tan2025}. 
One challenge of such peer-to-peer online spaces is the risk of becoming a space where antisocial proclivities are justified and neutralized, rather than acting as a therapeutic and prosocial resource \citep{sykes2017techniques, o2010content}. 
\subsubsection{Digital interventions} Online platforms and ESPs have developed extensive systems to detect, classify, report, and block CSAM. These efforts initially emphasized supply-side prevention through technical means such as classifiers, hash-matching and keyword detection \citep{westlake2012comparing, sujay2024comprehensive, google_csam_2025}. 
Over time, and in line with calls from civil society and researchers to strengthen secondary prevention \citep{quayle2020prevention}, attention has expanded toward demand-side measures that deter offending and promote help-seeking. Many sites, particularly hosting sites with user-generated content uploading capabilities and sites with search capability, have implemented relevant interventions \citep{IWF2024PornhubChatbot, protectchildren2025}. Warning messaging has been implemented by social media platforms including Instagram, Facebook, TikTok, and Snapchat, search engines including Microsoft Bing and Google Search, and on adult pornography sites such as Pornhub. In a scoping review, \citet{price2024review} report that such digital interventions can be effective particularly in deterring at-risk individuals or those still in early stages of exploration, but they noted a paucity of relevant studies. Upon detection of attempts to access known CSAM URLs or queries related to CSAM, platforms can serve warning messages and direct users to resources such as those detailed in section \ref{clinician} \citep{hunn2023how}. 

The content of warning messages across platforms was typically developed with the involvement of in-house trust and safety teams \cite{tyler2025new} drawing on available research and platform-specific expertise.  In particular, ESPs may lean on the emerging literature on encouraging engagement through  design, language and content choices for warning messages, chatbots, and other ESP-delivered interventions. Importantly, these have the potential to reach even those individuals not actively seeking help or referred to programs through justice system referrals \citep{smith2024chatbots, prichard2024effect, price2024review}. 

Early qualitative work underscored the importance of a non-judgmental and welcoming tone in digital prevention messaging, with participants emphasizing language that would be perceived as calm and hopeful would be more effective than communication based on fear and shame. \citep{henry2020designing}. However, these findings drew on insights from individuals who were already engaging with therapeutic services, and may not fully reflect deterrence and help seeking mechanisms for people actively seeking CSAM. Using controlled \textit{honeypot} websites designed to mimic illegal CSAM portals, Prichard and colleagues conducted a series of randomized experiments and found that men could be dissuaded from accessing the site through both therapeutic and deterrence themed warning messages  \citep{ prichard2022warning, prichard2022effects, prichard2024effect}. 
\citet{ociardha2025iatso} examined different warning message framings with both low- and high-risk individuals---classified based on self-reported past illegal or borderline behaviors (e.g., deliberate CSAM viewing, suspected underage sexual interactions) or stated attraction to online child-focused sexual behaviors. Study participants were asked to rank different hypothetical messages they would receive in the course of a search engine session on their perceived deterrence effect and effect on help seeking. Participants rated legality and harm-focused messages as most likely to deter further searching, while distress reducing self-efficacy increasing messages were seen as more likely to prompt help-seeking. Gain-framed variants—those emphasizing positive outcomes rather than punishment—improved help-seeking appeal. 

The honeypot experiments by Prichard and colleagues tested message framing in live settings, enabling ecological validity with naïve participants, but, for ethical reasons, focused on “barely legal” material as a proxy for CSAM and drew on modest sample sizes. In contrast, the work by \citet{ociardha2025iatso}, based directly on messaging as it would be encountered by Google Search users, engaged a much larger but more diverse sample but relied on perceived deterrence and help-seeking ratings rather than observed behavior. Together, these studies highlighted the need for large-scale field evaluations capable of assessing real-world behavioral impact.

Addressing these limitations, a large-scale evaluation examined the impact of an intervention implemented by Aylo (formerly MindGeek) on Pornhub, a major adult-content platform \citep{scanlan2024rethink}. In this intervention, users searching for CSAM-related terms were redirected to a warning page that stated such material was illegal and contained a link to a therapeutic support service, as well as an interactive chatbot called \textit{reThink}. The chatbot was designed to quickly provide information about available support services. 
The presence of the chatbot was associated with a statistically significant decrease in searches for CSAM, and a small fraction of users who interacted with the chatbot clicked through to the linked StopItNow UK website. More than 4 in 5 (82\%) of users who received the warning did not proceed to search again for a CSAM-related term in the session (although most did watch other legally-available videos and make other searches), representing a desistance in ongoing CSAM-seeking within the user's session \citep{scanlan2024rethink}. The study provided large-scale behavioral evidence of the potential for platform-delivered interventions to deter offending and promote engagement with support resources. 

\subsection{Google Search}
\label{searchdescription}
A ``Onebox,'' within Google Search, is a prominent Google-owned feature displayed at the top of the results page, appearing above the search results. Its primary purpose is to provide users with direct, concise, and relevant information or actionable features in response to specific queries, aiming to fulfill user information needs \citep{Zeiger2010EmergencyInfo, merritt_2022_suicide}. Other search engines have their own implementation of such a user interface \citep{t3_2013_bing}. 
 Google deploys the Search Onebox across a range of sensitive domains, including queries about public health (e.g., COVID-19), suicide and self-harm, child exploitation, and electoral processes. The Search Onebox is primarily surfaced through algorithmic means, including classifiers that are trained to identify different types of query intents. They can also be surfaced by keyword lists to supplement algorithmic triggering. The geographic specificity of Search Oneboxes varies - generally Google aims to serve bespoke local interventions, with relevant geographic resources and language. COVID-19 interventions are a good example of this, with the Search Onebox for COVID-19 pointing to local government health authorities \citep{google2020cor0170}.

\begin{figure}[!htbp]
    \centering
    \begin{subfigure}[b]{0.48\textwidth}
        \centering
        \includegraphics[width=\linewidth]{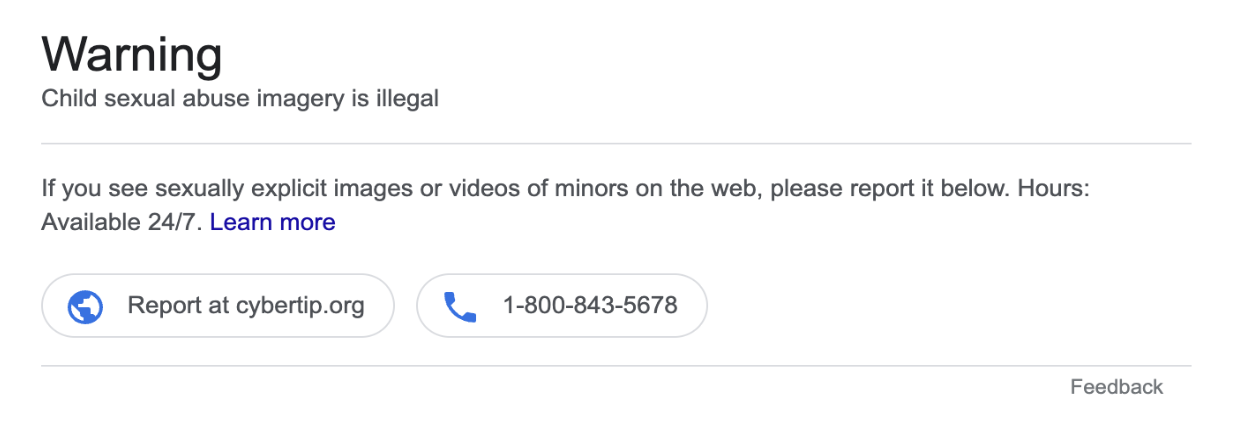}
        \caption{Onebox served to CSAM-seeking users prior to February 2025}
        \label{fig:oldonebox}
    \end{subfigure}
    \hfill 
    \begin{subfigure}[b]{0.48\textwidth}
        \centering

        \includegraphics[width=\linewidth]{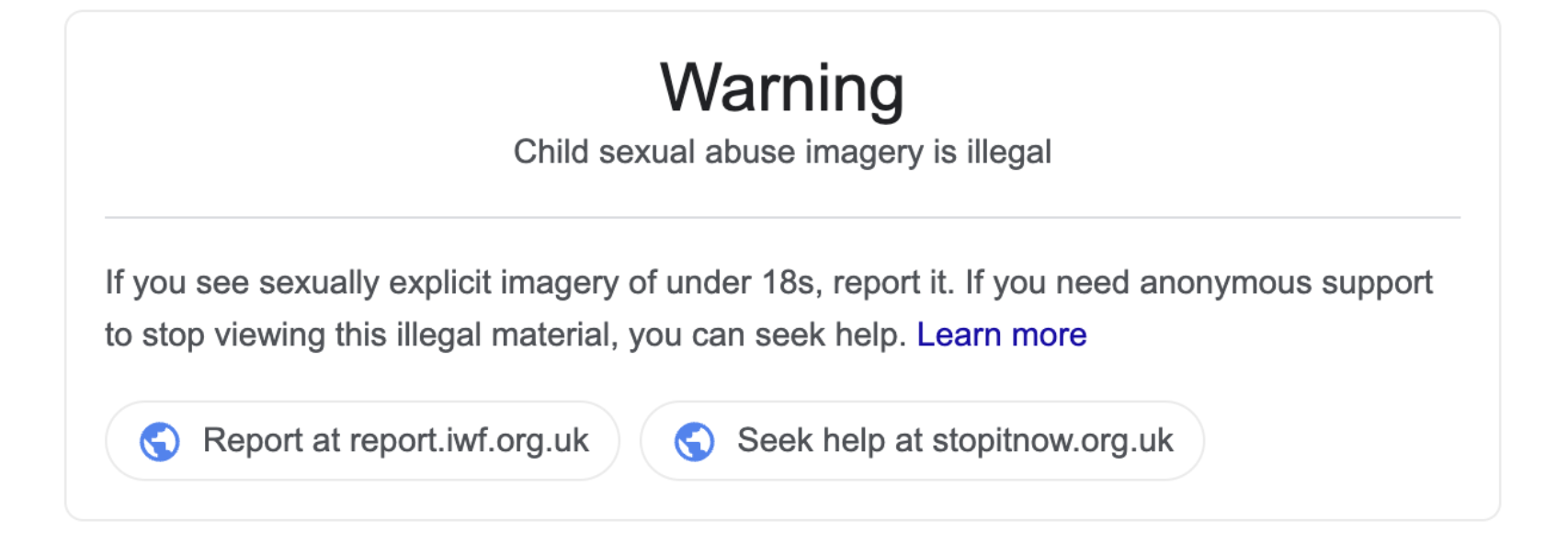}
        \caption{Onebox served to CSAM-seeking users located in England, Wales, Scotland, Northern Ireland prior to February 2025}
        \label{UKOB}
    \end{subfigure}
    
    \caption{Oneboxes prior to revision.}
    \label{fig:combined_onebox}
\end{figure}

The CSAM Onebox has been live for over a decade, launching in November, 2013 \citep{essers2013google} and covering 21 regions representing an aggregated population of 2.8 billion people \citep{WorldBankPopulation} by 2025.\footnote{This represents a ceiling estimate, as individuals in those countries may not have internet, may not use Google Search, or may use Google Search in a different language than is covered by the Onebox.} Figure \ref{fig:oldonebox} shows an example of the  messaging served at the top of the results page, which historically has focused on directing users to reporting hotlines.\footnote{The ``learn more'' hyperlink within the Onebox brought users to a more expansive list of resources, including in some cases, helpline resources.} 
Figure\ref{UKOB} shows how Google works to accommodate bespoke experiences, where UK-based users had a distinct version of the Onebox due to longstanding partnerships with the relevant organizations. Users receive this Onebox upon issuing a query that is classified as seeking CSAM, meaning the query includes some sort of sexually explicit component and a reference to a minor/child. The messaging represents part of a suite of digital interventions and protections designed to prevent 1) CSAM being indexed or served, 2) sexual imagery of adults being served in response to CSAM-seeking queries, 3) non abusive images of children being served for queries seeking adult pornography, and 4) other CSAM-related prompts being suggested \citep{canegallo2021efforts}.  From August 2024 to January 2025, on average, the CSAM Onebox was served in response to 20 million queries a month.\footnote{This includes false positives (e.g., the query ``child porn'' invokes the messaging, even though that query may be related to information seeking in response to news stories).} According to Routine Activity Theory \citep{cohen1979social} and the work conducted by \citet{prichard2024effect}, we expect this messaging to have at least a minor deterrent effect due to an increased perception of surveillance.

\section{This study}
In the current study, we analyze the effects of a change to the CSAM Onebox, wherein queries that would have previously returned messaging surrounding illegality and reporting resources, now return warning messaging oriented around the negative consequences of CSAM seeking and directing users to help-providing resources such as those detailed in section \ref{interventionsection}.

This study poses the following research questions:
\begin{enumerate}[label=\textbf{RQ\arabic*}, ref=RQ\arabic*, leftmargin=*]
    \item To what extent does the transition to therapeutic-oriented messaging outperform traditional reporting-focused prompts in reducing subsequent CSAM-related search queries? \label{rq1}
    
    \item Does the transition to harm-reduction messaging result in engagement with help-seeking services? \label{rq2}
    
    \item Does surfacing services within the Onebox result in a significant uptick in traffic to those services? \label{rq3}
\end{enumerate}

\section{Methods}
\label{methodology}
This study consists of an 1) expert workshop, where experts came together to ideate and approve specific designs and wording for a revised Onebox; 2) an evaluation of the effects of switching users to the new Onebox (the process of this switch is sometimes referred to as a ``launch''). 
\subsection{Expert Design Workshop}
Google convened a participatory design workshop in March of 2024, with a focus on re-imagining the CSAM Onebox.

Google participants included user experience researchers, product managers, policy analysts, and engineers. External participants invited for their expertise included 10 practitioners and academics (from Australia, Canada, Finland, Germany, Ireland, the United Kingdom, and the United States) with expertise in deterring CSAM seeking and providing therapeutic help to individuals with a sexual interest in children and other risk factors for child sexual abuse. Workshop participants were asked to focus on how to deter individuals who were seeking CSAM on Google Search, keeping in mind guidance around best user experience practices (e.g., the amount of text that can fit comfortably on a mobile screen, or in a drop down menu). 
The original message, with the exception of the UK, focused heavily on reporting to third parties such as the CyberTipline, IWF, and Safernet (Figure \ref{fig:oldonebox}).  As a result of expert feedback,  help-oriented messaging and resources were surfaced as the most prominent option, followed by victim support services and reporting services. This orientation acknowledged that people issuing CSAM-adjacent queries may be seeking help for themselves (e.g., victim-survivors looking to report their own content) or for loved ones. 
The messaging related to illegality and warnings was retained. The revised placement and language acknowledges that users issuing such queries are more likely to be served and helped by therapeutic resources, rather than information related to reporting illegal content. The language and placement of resources is also designed to reduce the stigma or fear that may prevent help-seeking.

\begin{figure*}[!htpb]
    \centering
    \begin{subfigure}[!htbp]{0.7\textwidth}   
        \centering
        \includegraphics[width=\linewidth]{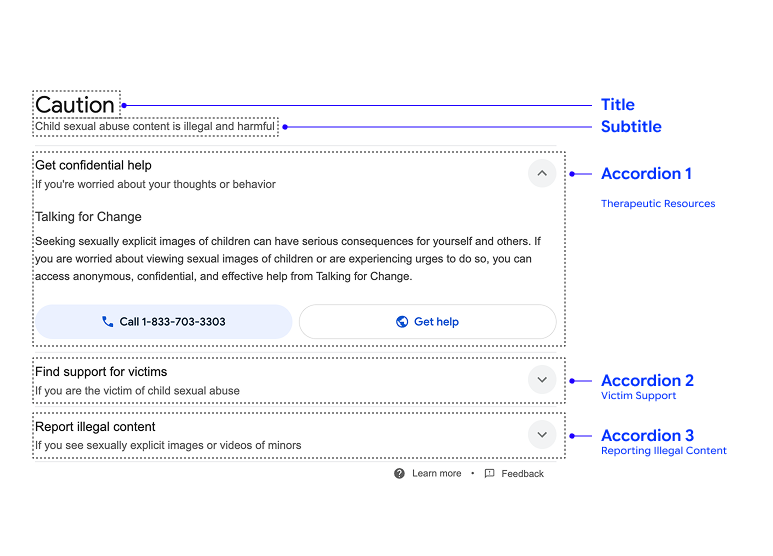}
        \caption{The default shows the help-serving language}
        \label{USOB}
    \end{subfigure}
    
    \par 
    \vspace{0.1cm} 
    
    \begin{subfigure}[t]{0.7\textwidth}   
        \centering
        \includegraphics[width=\linewidth]{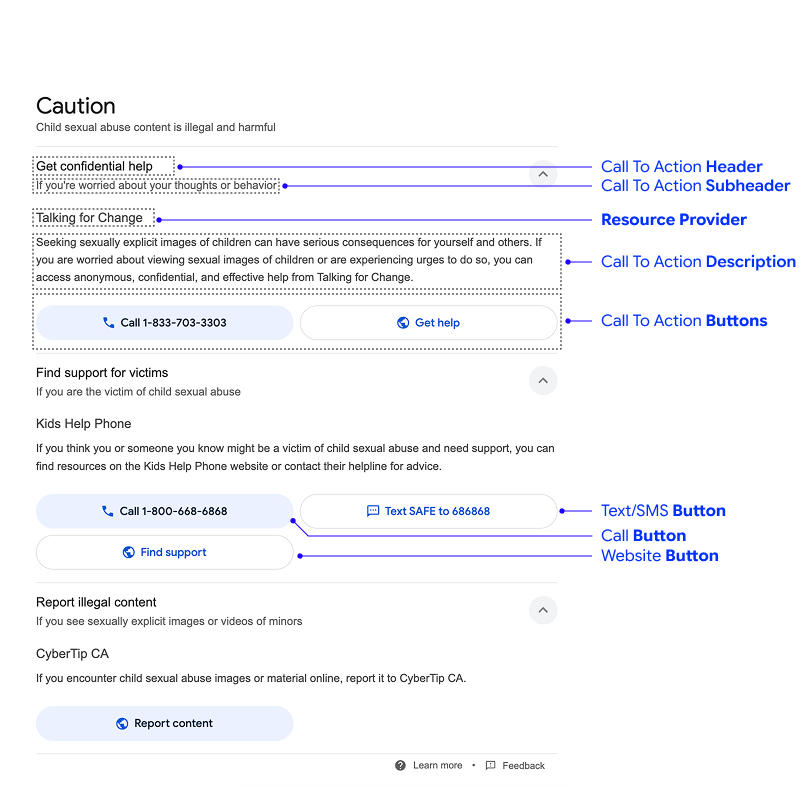}
        \caption{The full user experience when all drop downs are clicked}
        \label{fig:us}
    \end{subfigure}
    
    \caption{Interface states of the redesigned Google Search CSAM Onebox.}
    \end{figure*}

The prototype then went through internal review at Google, and the final version (Figure \ref{fig:us}) was approved by all workshop participants. 

\subsection{Launch approach and region coverage}

\begin{table*}[!htbp]
    \centering
    \small
    \begin{threeparttable}
        \caption{Treated country/language pairs and available help modalities linked within the Oneboxes in Wave 1. Only therapeutic services shown.}
        \label{countries} 
        \begin{tabular}{lllcccc}
            \toprule
            Treated Countries & Treated & Help Org & \multicolumn{4}{c}{Help Modalities Offered\tnote{a}} \\ 
             & language(s) & & Chat & Call & Website & SMS \\
            \midrule
            Australia & English & Stop It Now! AUS & n & y & y & n \\
            Belgium & French/Dutch & SeOS (fr)/ Stop it Now! BE & n & y & y & n \\
            Canada & French/English & Parler pour Changer/Talking for Change & y & y & y & n \\
            France & French & Stop Device & n & y & y & n \\
            Germany & German & Kein Täter werden & y & y & y & n \\ 
            Republic of Ireland & English & Stop It Now! IE & y & y & y & n \\
            Netherlands & Dutch & Stop It Now! NL & y & y & y & n \\
            New Zealand & English & Safe Network & n & n & y & n \\
            US & English & MOORE & n & y & y & n \\ 
            United Kingdom & English & Stop It Now! UK & y & y & y & n \\
            Republic of Korea & Korean & \textit{None} & - & - & - & - \\
            \bottomrule
        \end{tabular}
        \begin{tablenotes}
            \item[a] Modalities offered within the Onebox interface at launch.
        \end{tablenotes}
    \end{threeparttable} 
\end{table*}
As described in section \ref{searchdescription}, Google works to partner with local in-country resources for Oneboxes. Accordingly, eleven regions in total  switched to a new version of the warning on February 12, 2025. The regions, help organizations, and modalities offered within the Onebox itself can be seen in Table \ref{countries}.  While the overall text was translated identically across countries, the users in each country were directed to the relevant local resource. 

A second wave of changes happened on February 24, 2025 for the remaining 10 countries with Oneboxes (Argentina, Brazil, Colombia, Spain, India, Italy, Mexico, Peru, Portugal, Thailand). While some do have relevant services (such as PrevenSi in Spain and Red PaPaz’s Te Guío in Colombia), Google has not yet established partnerships. Thus, rather than a significant change of both wording and proffered third party resources, these countries received more subtle updates to phrasing and language, and resources surfaced included reporting hotlines and victim-support resources. We did explore analyzing the effect of this minimal change, but lacked an appropriate comparison group and thus this change is out of scope for this study.

\subsection{Dataset}

Our dataset comprised a country-stratified random sample of anonymized search sessions where users issued at least one CSAM-seeking search query. A search session consists of an uninterrupted or mostly continuous flow or queries and interactions. 
A session may last anywhere from several minutes to a whole day depending on various factors.
Each session also includes high-level metadata such as the country, user's set display language, and date. We collected approximately three weeks of data prior to the launch of the new CSAM search messaging (February 12) and 14 weeks post-launch, specifically from January 17, 2025, through May 20, 2025. 

The data used in this analysis is managed according to Google's data policies to protect user privacy \citep{google_priv-dj}. This means that the data undergoes regular, multi-stage de-identification and pseudonymization processes \citep{google_anon-re}. These privacy-preserving anonymization measures (like the removal or obfuscation of potentially identifying information or \textit{PII}) are applied at regular interval and some of these techniques have been open sourced \citep{Forget2019-xz}. All use of data sourced from such logs are subject to an internal ethics, legal and privacy review process and approval to meet Google's privacy standards.

Furthermore, our data collection time period window was constrained by two factors:
\begin{enumerate}
    \item \textbf{Sensitive Data Retention Policy:} In accordance with GDPR storage limitation principle, Google only retain highly granular sensitive user data for as long as there is a legitimate business need. In line with this policy, we could not track user data that fell outside this data retention period. This data retention period, for our particular analysis, was a 90 day trailing window from the day of data collection.
    \item \textbf{Identifying an appropriate convenience comparison group of countries:} We could only arrive at a consensus for an appropriate group of convenience comparison countries by 17 April 2025, which meant we could only collect data as far back as 17 January 2025.
\end{enumerate} 

\subsubsection{Country and Language Selection}
\paragraph{Experimental/Treated group}
Table \ref{countries} shows the regions considered for inclusion in the experimental group. We \textit{a priori} excluded two regions from the main analyses in the interest of comparability and parity in change and experience across the experimental group. Google has been unable to identify an appropriate help-seeking partner organization in South Korea, so the main change in South Korea was the surfacing of victim-support resources in addition to reporting hotlines. The United Kingdom (England, Wales, Scotland, Northern Ireland) 
has had for several years now, a distinct version of the warning message (Figure \ref{UKOB}). This messaging had elements of both the original messaging (the first resource is a reporting option, the first half of the messaging is focused on reporting) and the new messaging (urging users to seek anonymous help and providing the link to Stop It Now!). Excluding these two regions left us with nine countries in our main experimental group. 

Because the UK messaging already had an existing button to the Stop It Now! website, we retained the UK data to conduct a simple pre/post analysis (calculating number of sessions where a user clicked on the resource/number of overall sessions in the time period).

\paragraph{Convenience Comparison Group}
We selected a convenience comparison group consisting of seven country/language pairs: Austria/German, Denmark/Danish, Sweden/Swedish, Switzerland/German, Norway/Norwegian, Iceland/English, and Luxembourg/English. We prioritized ecological similarity over geographical proximity. Specifically, these countries were selected because they match the treatment group on the following parameters to ensure valid comparison:
\begin{enumerate}
    \item \textbf{Comparable Internet Penetration Rates:} All selected countries exhibit a high internet penetration rate of over 93\% \citep{ITU2024-ez}. While we cannot closely match for ``propensity to seek'' CSAM material between our treatment and convenience comparison regions, we aimed to select regions for our convenience comparison group that have similar digital maturity to our treatment group to ensure that the baseline rate of ``opportunity to seek'' CSAM material is similar between the two groups. This is important because online offending requires the convergence of both a motivated offender and a target \citep{cohen1979social}.
    \item \textbf{Comparable Search Ecosystem:} In all of our selected countries, Google is the search engine provider of choice for information seeking with the percentage of total search queries processed by Google compared to the total volume of searches conducted in the country being above 80\% \citep{statcounter_austria_2025, statcounter_denmark_2025, statcounter_sweden_2025, statcounter_switzerland_2025, statcounter_norway_2025, statcounter_iceland_2025, statcounter_luxembourg_2025}. This allows for necessary ecological parity between the two groups as it allows us to capture the behavior of the general populations as opposed to a biased sub-segment of users and ensures that the "search-to-intervention" user journey is comparable.
    \item \textbf{Comparable Legal Frameworks:} All selected countries are signatories to the Lanzarote Convention or equivalent protocols \citep{Council-of-Europe2007-vi}, establishing a shared deterrence context where CSAM seeking is uniformly criminalized and stigmatized. This ensures parity in terms of punishment costs associated with CSAM seeking with the treatment group.
    \item \textbf{Comparable Therapeutic Referral Infrastructure:} Each of our selected convenience comparison regions possess a standardized and anonymous therapeutic intervention ecosystem that allows for parity of potential destination in the ``search-to-intervention'' user journey. So even in the absence of a Onebox in these regions, users would still find valid therapeutic resources on the search engine result page.
\end{enumerate}

We then conducted a country-stratified random sample of CSAM-seeking search sessions from these regions, making all language queries eligible for the sample. Table \ref{samplings} shows the regions and languages included in each of the difference-in-differences analyses used to answer \ref{rq1}.
\begin{table*}[!htbp]
\footnotesize
\centering
\caption{Regions and languages included in each analysis}
\begin{tabular}{llcccccc} 
    \hline
   \textbf{Region}&\textbf{Language}& \multicolumn{6}{c}{\underline{\textbf{Analysis}}}  \\
     && \multicolumn{2}{c}{\textbf{1}} & \multicolumn{2}{c}{\textbf{1a}} & \multicolumn{2}{c}{\textbf{1b}} \\
    \cline{3-4}\cline{5-6}\cline{7-8} 
    \multicolumn{2}{c}{} & experimental & convenience  & experimental & convenience & experimental & convenience \\
      \multicolumn{2}{c}{} &  & comparison &  & comparison &  & comparison \\
    \hline
    Australia & English & x & & & & & \\
    Austria & German & & x & & x & & \\
    Belgium & French & x & & x & & x & \\
    Belgium & Dutch & x & & x & & x & \\
    Belgium & English & & & & & & x \\
    Canada & English & x & & & & & \\
    Denmark & Danish & & x & & x & & \\
    France & French & x & & x & & x & \\
    France & English & & & & & & x \\
    Germany & German & x & & x & & x & \\
    Germany & English & & & & & & x \\
    Ireland & English & x & & x & & & \\ 
    Netherlands & Dutch & x & & x & & x & \\
    Netherlands & English & & & & & & x \\
    New Zealand & English & x & & & & & \\
    Norway & Norwegian & & x & & x & & \\
    Sweden & Swedish & & x & & x & & \\
    Switzerland & German & & x & & x & & \\
    US & English & x & & & & x & \\ 
    US & Spanish & & & & & & x \\
    \bottomrule
\end{tabular}
\label{samplings}
\end{table*}
\subsubsection{Sampling Method}
Raw user activity data for search sessions are stored in a distributed file system architecture \citep{Ghemawat2003-ge}, where dataset is partitioned into thousands of distinct file segments (`shards') to enable parallel processing \citep{Melnik2011-mb}. To analyze the data in a memory efficient manner, we used a file-based sampling strategy where we randomly selected 10\% of these file segments for analysis \citep{Samwel2018-py}. Since user activity data are evenly and randomly distributed across these files, this method provided for a representative random sample of the larger population without having to process every single record in a memory intensive fashion.

For region/language pairs in both the treatment and the convenience comparison group, we sampled a representative random sample of 10\% of search sessions where a user had issued a CSAM-seeking query on Google Search. The groups differed only in their exposure to the CSAM Onebox. The treatment group encountered the Onebox (both before and after the launch of the new design \footnote{The treatment group users experience the old Onebox design in the pre-intervention period and the new Onebox design in the post-intervention period.}), while the convenience comparison group received the standard Search Engine Result Page  with CSAM-related results filtered out throughout the data collection period.

Where region/language pairs resulted in fewer than 100 observations per day, we boosted our sample size. We eliminated country/language combination samples in both the convenience comparison and experimental groups that fell below our requisite minimum sample size of 100 samples per day over the entire 17 week period. This approach allowed us to retain almost all experimental country/language conditions, with the exception of Canadian French, primarily spoken in Quebec. It also led us to eliminate two low-sample convenience comparison regions (e.g., Luxembourg, Iceland), leaving us with five remaining regions. \footnote{We expanded the parameters for Luxembourg and Iceland to include sessions in any language, but still felt short.}

For the main analysis (Table \ref{samplings}, column 1a), our experimental group consisted of sessions in treated countries and treated languages ($n = 696,653$), with our convenience comparison group consisting of untreated countries and untreated languages ($n=371,673$). We considered that the majority of our convenience comparison group were European countries, so we conducted two robustness analyses. For one, we filtered out non-European sessions from the first analysis to compare just European to other European countries (total $n=710,428$, Table \ref{samplings}, column 1b). For the other (Table \ref{samplings}, column 1c), we narrowed down our experimental group from the first analysis to include only those countries which also had at least one untreated language with sufficient sample size ($n=436,525$), and our convenience comparison ($n=395,532$) group was the sessions in those untreated languages (e.g., Spanish in the United States).

\subsubsection{Outcome variables}
Our primary outcome variable was whether a user, in a single session, issued a new query classified/labeled as CSAM seeking after their first CSAM seeking query (dichotomized, 0=no reissue, 1 = reissue). We thus aggregated, by week, to examine the change in the follow on CSAM query rate attributable to the new Onebox experience. 

\subsection{Helpline Data}
In some cases, helplines shared aggregated traffic data  (typically in the form of counts per day). Where possible, we filtered further, to either only that data directly attributable to Google broadly(both via the organic results or the Onebox), or via the Onebox specifically. We manually applied a preliminary restriction to the data to include just data from the beginning of 2025 to May 20th.
\subsection{Analyses}
Difference-in-differences (DID) methods are often leveraged to examine the causal effects of policy changes or natural experiments by comparing the change in the outcome over time in a treated group with the change in the convenience comparison group, under an assumption of parallel trends (i.e., in the absence of the intervention, outcomes in the
intervention group and the convenience comparison group would have remained parallel over time).  We used the DID package within R \citep{callaway2021getting} for our primary analyses. This package is designed for robust difference in differences analysis. 
To address \ref{rq1}, we conducted one main analysis and two robustness analyses: 
\begin{itemize}
    \item Analysis 1: Effect of user interface change \& introduction of help seeking (treated vs convenience comparison group of untreated languages in convenience comparison countries) \label{analysis1}
    \item Analysis 1a: Robustness analysis (European treated countries/languages vs convenience comparison of untreated European countries) \label{analysis1a}
    \item Analysis 1b: Robustness analysis (Treated vs convenience comparison of untreated languages in treated countries) \label{analysis1b}

    \end{itemize}

Within the Onebox itself, depending on the service, there are buttons that direct the viewer to the website of the helpline, and/or to a phone number to call directly. To answer \ref{rq2}, we calculated the percentage of sessions within our experimental sample ($n=696,653$) that clicked on any service button. We also leveraged the fact that the UK had a version of the new Onebox prior to launch to look at the rate of sessions clicking on a helpline button in the month immediately prior to and following the launch date in our UK data. We artificially abridged the data to reduce the likelihood of data contamination effects.

\begin{itemize}
    \item Analysis 2: Effect of user interface change \& introduction of help seeking on percentage of sessions with at least one click on help seeking resources.\label{analysis2}

    \end{itemize}

Finally, helplines themselves often retain data on how users reach their websites. This can provide convergent evidence to our data, and answer \ref{rq3}. Some helplines (Talking for Change, Kein Täter werden, Stop It Now! Aus, Stop It Now! IE, Stop It Now! UK) chose or were able to share anonymized traffic volume data, aggregated to the day level.\footnote{Help Wanted/MOORE in the United States offered data, but due to a switch in web providers in early 2025, traffic data was unreliable and thus unsuitable for inclusion. Dispositif STOP in France provided weekly-level data, available in table form in Appendix \ref{tab:visitor_data}.} Depending on the settings and data analytics software, helpline data had varying levels of filtering and annotation. In Australia, Ireland, and the UK we used the data attributable specifically to the Onebox (before and after the revision). In other cases, we examined any data attributable to Google, which includes Google Search and the Onebox, but would exclude traffic coming directly from ad campaigns on social media (Germany). Finally, for Canada, we had to include all traffic data, regardless of how the user reached the help service web page. Thus, \hyperref[analysis3]{Analysis 3} used an interrupted time series design \citep{kontopantelis2015regression} to assess whether the launch led to a significant increase in helpline traffic.
\begin{itemize}
\item Analysis 3: Interrupted time series analysis of effect of user interface change \& introduction of help seeking on helpline traffic \label{analysis3}
\end{itemize}

\section{Results}
\label{results}


\subsection{Analysis 1: Effect of User Interface Change and Introduction of Help Seeking}
\subsubsection{Analysis 1}
\begin{figure*}[!htpb]  
\centering
\includegraphics[scale=.7]{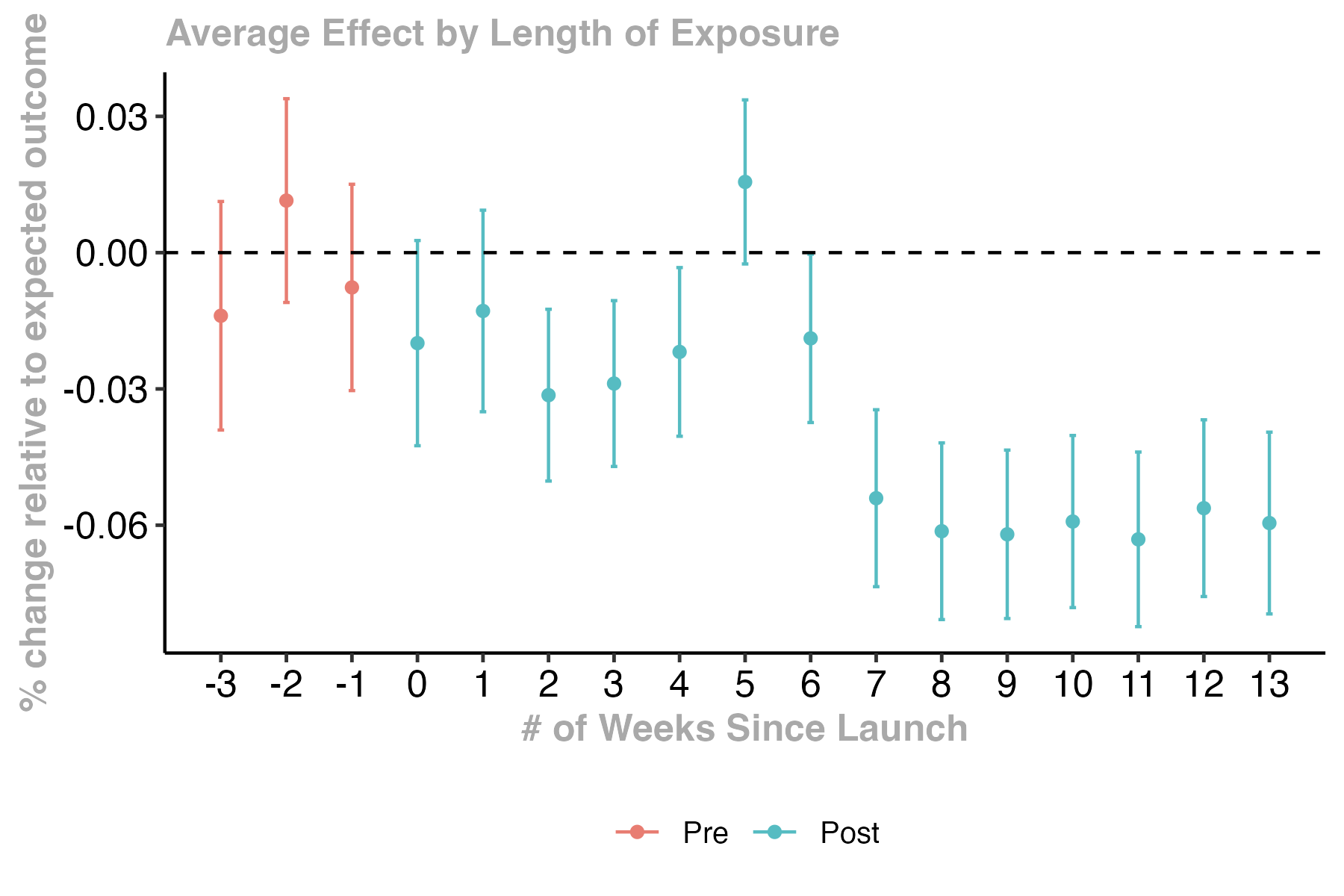}
\caption{Treatment effect over time. Week 0 represents data from Feb 12 (launch) to Feb 18}
\label{did}
\end{figure*}

Our primary difference in differences analysis compared rates of ongoing CSAM-seeking on Search between our five convenience comparison countries and our experimental group, before and after launch of the new Onebox. That is, we analyzed whether sessions go on to issue a new CSAM-seeking query after their initial one.
 Our parallel trends assumption was not violated ($p=0.33$). We found that the new  messaging resulted in a 3.8 percentage point  ($SE=0.006, 95\%CI=-0.050, -0.026, p<.05$) decrease in re-issuing of CSAM-seeking queries compared to their predicted outcome if there had been no change. The current analysis, as seen in Figure \ref{did}, showed that the drop in re-issuance reached significance approximately two weeks after the launch. For week 5 and 6, we observed an anomaly in the data, which manual review attributed to external news events (see discussion). The negative effect emerged strongly and seemed to stabilize 7 weeks after launch.

\subsection{Robustness Checks}
\subsubsection{Analysis 1a: Only European Countries}

\begin{figure*}[!htpb]  
\centering
\includegraphics[scale=.7]{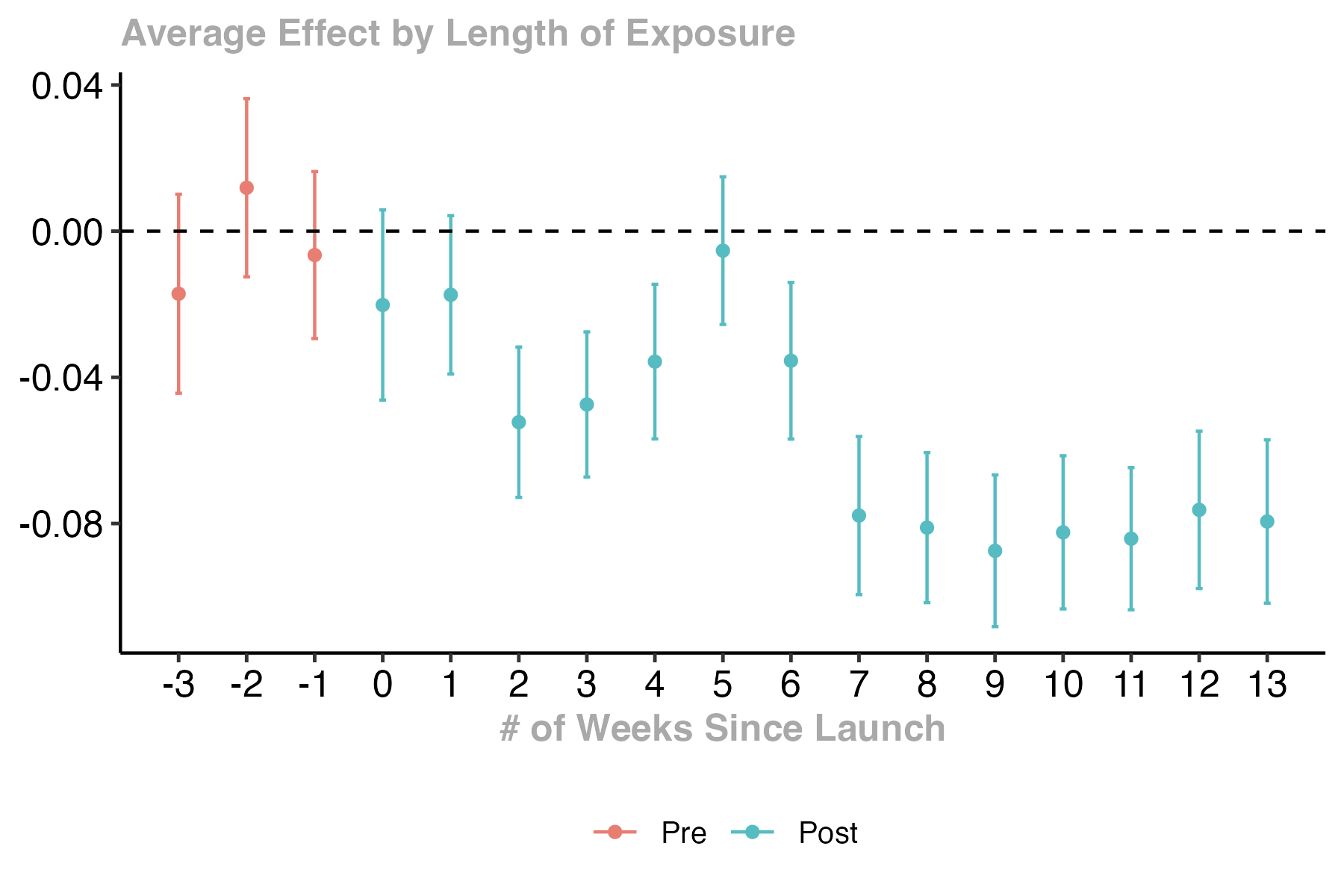}
\caption{European countries difference-in-differences analysis}
\label{europe}
\end{figure*}

We narrowed down our experimental group to include only European countries, in the interest of even more precise comparability in our convenience comparison and experimental groups (as all of the countries in the former group were European). The parallel trends assumption still held ($p=0.33$). Results were slightly strengthened, with a  5.6 percentage point decrease in re-issuing CSAM-seeking queries ($SE=0.007, 95\%CI=-0.069, -0.043, p<.05$), seen in Figure \ref{europe}.




\subsubsection{Analysis 1b: Treated vs untreated languages in experimental countries}
We looked at the subsample of five treated countries that had sufficient sample sizes in an untreated language (e.g., Spanish in the United States).  Our parallel trends assumption was violated ($p<.05$), so we were unable to proceed. 

\subsection{Analysis 2: Percentage of Sessions with Clicks on Therapeutic Services}
In addition to measuring re-issuing of CSAM queries, we assessed the percentage of sessions which included a click on the help-providing resources. As noted in Table \ref{countries}, some countries have buttons to click for a direct call, while others direct users only to the website. Across all modalities and treated country/languages, 0.73\% of sessions contained a click on any of the hyperlinked buttons to the help-providing resources. 
\subsubsection{UK Pre/Post}
The UK had a version of the CSAM Onebox that included a ``Seek help'' button for Stop It Now (see Figure \ref{UKOB}), making it unsuitable for inclusion in the main analyses but creating an opportunity to look at the impact of the amended wording and user interface. Analyzing just the sessions in the month immediately preceding and following the launch, the percentage of sessions that included a click onto the resources rose from 0.5\% before the launch, to 1.7\% following. 

\subsection{Analysis 3: Helpline Data}
\label{helpline}

\begin{figure}[htbp!]
    \centering
    \includegraphics[width=0.8\linewidth]{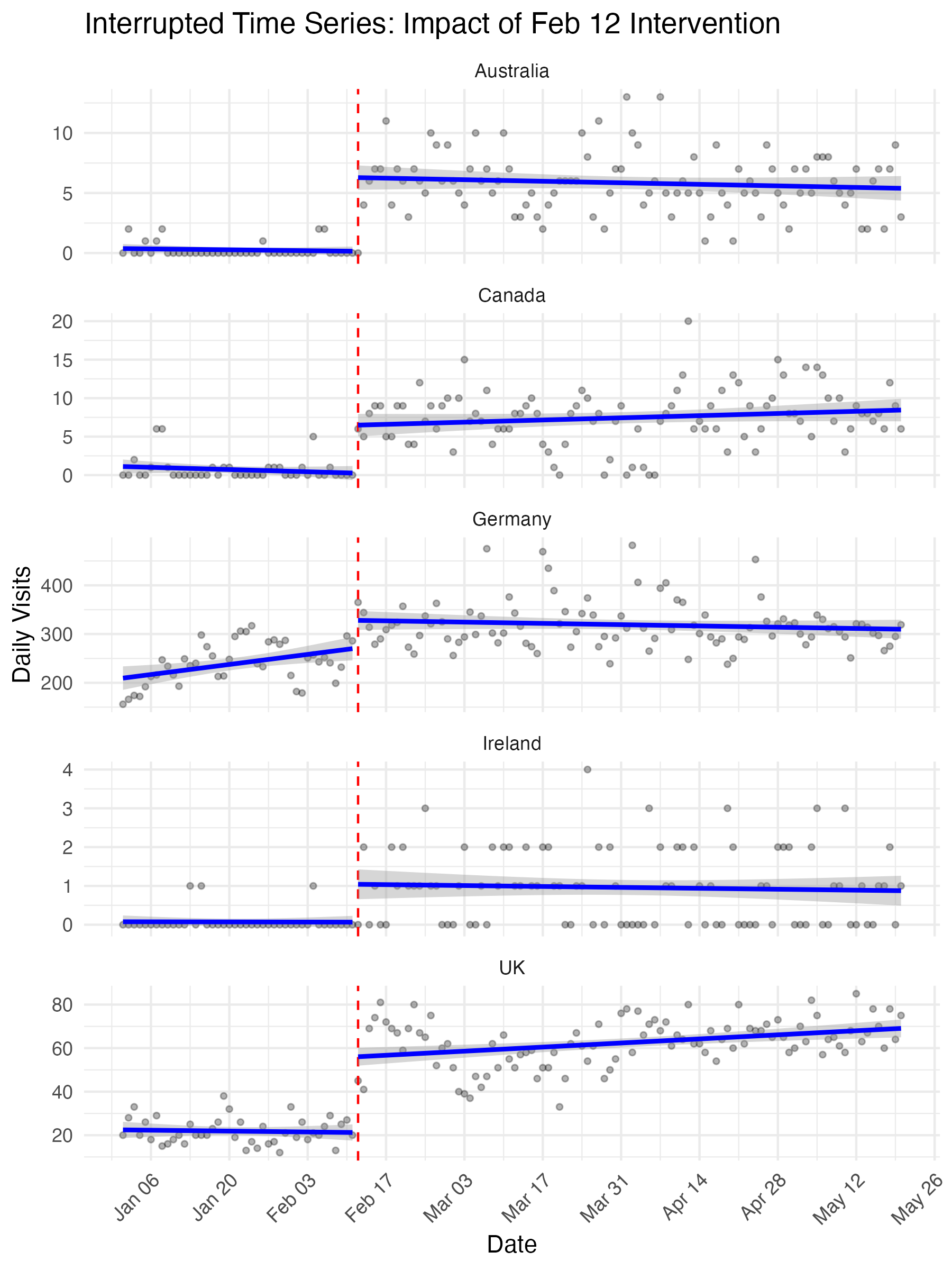}
    \caption{Traffic patterns from helplines, by region}
    \label{fig:Helpline data traffic changes}
\end{figure}
The interrupted time series analysis revealed an immediate and significant jump in helpline traffic attributable to the launch, across all five regions. Depending on the country, the launch resulted in between one additional visitor to the website per day (Ireland) up to 58 more per day in Germany, although the large standard error and analysis of the ``pre'' data suggests that concurrent efforts such as social media marketing campaigns may have introduced noise into the model. Full model results can be seen in Table \ref{tab:its}.

\begin{table}[!bpht]
\centering
\caption{Interrupted Time Series Analysis: Immediate Impact and Trend Changes by Region}
\label{tab:its}
\begin{tabular}{lccccc}
\toprule
 & \multicolumn{2}{c}{\textbf{Immediate Change}} & & \multicolumn{2}{c}{\textbf{Change in Trend}} \\
 & \multicolumn{2}{c}{(Intervention)} & & \multicolumn{2}{c}{(Time After Intervention)} \\
\cmidrule{2-3} \cmidrule{5-6}
\textbf{Region} & \textbf{Coeff.} & \textbf{SE} & & \textbf{Coeff.} & \textbf{SE} \\
\midrule
Australia & 5.96$^{***}$ & (0.94) & & -0.01 & (0.03) \\
Canada    & 6.22$^{***}$ & (1.63) & & 0.04  & (0.06) \\
Germany   & 57.90$^{*}$  & (25.79) & & -1.73 & (0.95) \\
Ireland   & 0.98$^{**}$  & (0.31) & & -0.00 & (0.01) \\
UK        & 33.35$^{***}$ & (5.02) & & 0.14  & (0.18) \\
\bottomrule
\multicolumn{6}{l}{\footnotesize $^{***}p < 0.001$, $^{**}p < 0.01$, $^{*}p < 0.05$} \\
\end{tabular}
\end{table}


\section{Discussion}
\label{discussion}
In this study, we investigated whether updated language, including direct links to help-providing services and delivered at the top of a search engine results page, resulted in a) a reduction in on-seeking on CSAM on Search relative to what would have been expected without the update b) some number of click-throughs to the relevant help-providing services, and c) an uptick in web traffic as reflected in the help-providing services data.

\subsection{Reduction in rate of follow-on CSAM-seeking queries}
We found that reorienting language and directing seekers of CSAM to help-providing, therapeutic resources, results in a significant, albeit small, reduction in re-issuing of CSAM-seeking queries within the same session. Importantly, this compares the results of the new box to what would have happened under status quo, so these results indicate that the new CSAM search messaging creates deterrent effects \textit{above and beyond} those of the original messaging. Google has had a warning since 2013, and while this study doesn’t evaluate the effectiveness of that message, the literature suggests that it would have had a dampening effect on search volume prior to the commencement of this study \citep{steel2015web}. 
We conclude that the reduction in issuing follow-on CSAM-seeking queries in our experimental group is a positive sign of deterred CSAM-seeking. Previous studies have also evaluated the specific effects of tone and approach \citep{ociardha2025iatso, henry2020designing}, varying emphasis to focus more or less on illegality, de-stigmatizing of help-seeking behavior, and potential consequences of behavior. Similarly and critically, our study did not evaluate the effect of having this feature at all, but rather the effect of a change in the feature from focused entirely on illegality and directing towards reporting resources, towards language and user interface features that together highlight illegality, harm, and consequences, \textit{and} signpost people towards effective, confidential, and anonymous help.

This finding also supports prior research regarding deterrence of CSAM seeking \citep{prichard2024effect}. We caution against overinterpretation of the findings, and echo longstanding calls for a socio-ecological approach to preventing child sexual abuse \citep{letourneau2014need}, including CSAM. Improving our approach to CSAM has become increasingly urgent as AI tools enable the creation of synthetic CSAM \citep{grossman2025ai}, and CSAM proliferates across the internet. 
\subsubsection{Post hoc review of week 5 anomaly}
 Manual review of trends revealed the possibility that a data contamination effect impacted our convenience comparison data in particular. Two events occurred on March 21, which corresponded with the beginning of that week, that may have driven an uptick in relevant traffic in convenience comparison countries. In Denmark, a former Danish politician and prominent businessman was arrested for CSAM. A few days later, the identity of the man was revealed, continuing to drive interest in our control countries \citep{cphpost_2025_exminister}. For the relevant week (March 19-March 25), according to Google Trends four of the top five rising search queries in Denmark were related to this incident \citep{google_trends_2025_online}.   Additionally, in Austria, there was a police crackdown on individuals calling themselves ``pedo-hunters'' and committing hate crimes against gay men \citep{theinternational_2025_arrested}.These newsworthy events likely introduced an influx of Search sessions likely to resemble the exact pattern we looked for in our experimental countries (sessions with a single CSAM-related query) , thus artificially deflating the difference between the experimental and the convenience comparison groups. 
\subsection{Interest in services}
Aggregated across the nine experimental regions included in Analysis 1, a small proportion of sessions (0.73\%) included a click onto one of the therapeutic resource buttons offered directly in the messaging. Considering our ``treated'' group of almost 700,000 sessions represented just a sampling of the universe of relevant sessions, that still means 5,000 sessions clicked into the resources directly from the Onebox. Available data provided by the help services point to a consistent uptick in traffic, as seen in section \ref{helpline}. While together we believe these numbers present a fuller picture than either alone, we note that both of these numbers exclude any phone-based interactions, thus our estimates still represent a likely underestimation of the effect.

We also acknowledge this number is low, compared to ``click through rates'' for other issues such as suicide \citep{onie2023suicide}. The language was crafted explicitly and intentionally to highlight those elements which could encourage help-seeking, however barriers like those listed in section \ref{background} still persist.

\subsection{Help services data}
An interrupted time series analysis of data provided by a subset of the help services directed to within the messaging showed a significant uptick in traffic as an immediate result of the intervention. This provides convergent evidence from an independent data source that our new messaging and the prominent placement of the services helped drive users to explore the help services.  

\subsection{Implications}
Our study suggests that a non-trivial number of individuals may be deterred and even seek out therapeutic services for their behavior and urges. Taken together, our findings suggest that digital interventions can play a role in CSAM demand-side prevention efforts, and that the language and user experience of such digital interventions should be carefully designed to emphasize relevant resources, and key features of those resources that can encourage engagement (confidentiality and anonymity). 

The help-serving language served three main purposes 1) communicating that CSAM is illegal, harmful, and has serious consequences 2) there is anonymous, confidential help available and 3) identifying and providing an easy way to access help directly. We also underline the value of how this messaging was conveyed. Researchers have shown that users trust Google to rank relevant information appropriately. By surfacing this messaging at the top of the results, and including the critical information of the services (such as confidentiality and website information) directly within the feature, the services may benefit not only from greater awareness, but also from users' high trust in Google and Google's ranking ability \citep{pan2007google}.

We situate our findings within the larger digital intervention landscape documented by \citet{prichard2024effect}, wherein Google Search returns this messaging at the top of a results set of educational and news results. Taking an even broader perspective, it is critical to consider the larger environment in which this message was being delivered. The challenges related to CSAM are only increasing as AI-generated CSAM floods the web \citep{IWF2024AI}, platforms move towards privacy-forward encryption \citep{teunissen2022child}, and perpetrators take advantage of jurisdictional limitations to carry out scaled financial sextortion schemes \citep{cretu2024unravelling}. An implication of this work is the critical need for appropriate evaluation and quantification of CSAM-prevention efforts that can help bolster support for sustained and sustainable funding. Politically, while the costs of incarcerating convicted perpetrators are accepted as necessary, services aimed at primary and secondary prevention are faced with the challenge of proving the benefits of a counterfactual (proving that abuse was averted). For example, one early program in the UK (i-SOTP) was discontinued due to cost, despite early encouraging evidence that the program had positive effects on individual psychosocial risk factors \citep{middleton2009does}. 

While services related to mental health and child sexual abuse reporting are widespread and global, services oriented at helping individuals manage inappropriate sexual urges are less common. It is critical that users seeking help are met with the appropriate resources, particularly because early failure or disappointment may lead to long-term disengagement. This study showed that at least some individuals seeking CSAM can be successfully redirected to therapeutic services. Thus, growing these programs to additional countries and languages should be a significant policy priority. 

\subsection{Strengths and Limitations}
\label{Limitations}
This study has a number of strengths, including objective metrics of deterrence and interest in therapeutic resources, multi-country analyses, and a robust amount of data. The ``in the wild'' nature of this study lends itself both to specific strengths (that is, external validity of findings), and notable limitations associated with constraints of measuring intervention effects \citep{chamberlain}. 

Our parallel trends in our difference-in-difference analysis relies on the available data, which was three weeks of data prior to the experimental change. More pre-change weeks of data would provide more confidence in the suitability of the convenience comparison group. The launch approach eliminated our ability to examine the effects of messaging alone, as seen in wave two countries, as we were unable to find an acceptable convenience comparison group. 

In our main analyses, the prerequisites for the difference-in-difference analysis were met, but we did encounter a contamination effect (i.e., news event in Denmark). A broad recommendation of this work is for ESPs to adopt warning messaging, a more specific recommendation is to do so in a way that supports more robust methodologies, such as a randomized controlled trial. 

We included helpline data where it was available as much as possible. Data not represented includes those from regions/helplines who did not respond to outreach (e.g., SeOS, Stop It Now! Belgium), where there were quality issues with the initial launch (Safe Network), and where provider-initiated platform migrations led to missing data during the time of evaluation (MOORE). In our analyses, we were also limited to the granularity of data offered, which meant that in some cases we were able to only evaluate change in traffic attributable to the Onebox (Stop it Now! Australia, Stop it Now! Ireland, Stop It Now! UK), in others we had to look at broader traffic patterns (Kein Täter werden, Talking for Change/Parler pour Changer), and in the case of Dispositif STOP, in a weekly format as seen in Supplemental Material in Table A1. 

Together, our click through rates and the helpline-provided traffic data point to a significant, albeit small, interest in seeking help. Importantly, both measures exclude those individuals who use their phones to call or text directly (rather than clicking on the button within the messaging). Additionally, we observe that not clicking on these resources does not mean a full scale rejection of the premise, but that it may take individuals time to actually explore these services. Delays in seeking treatment are a well-documented phenomenon in a range of areas \citep{laumann2009population, ten2013lifetime}.  
 
This experiment was only tracking behaviors within a single session. It is possible that the deterrence observed is not sustained if considering users longitudinally, that is, it disrupted activity for a limited period of time but did not result in long-term behavioral change. 
We measured platform-specific, within-session, desistence of CSAM-seeking. It is possible the new treatment resulted in displacement, causing users to seek CSAM in other places, such as other search engines, or on the dark web. 

\subsubsection{Targeted population}

This intervention  only reached those individuals looking for this kind of content on Google Search (the open web). We are thus missing savvier users who may be using the dark web and/or more peer-to-peer modes of transmission. Additionally, classifiers are not 100\% effective in detecting relevant queries, so a subset of relevant Google Search queries missed by the classifiers would not have been served the Onebox. Although this means it may be limited in efficacy for people in advanced stages of offending, nevertheless Google's intervention can act as a bastion against initial usage and further transiton to TOR or other modalities \citep{steel2022technical}. 

Finally, use of Virtual Private Networks (VPNs) may introduce a specific form of measurement bias. VPNs are tools that allow users to obfuscate their real IP address as VPN providers serve their website traffic from a different IP address \citep{berger2006analysis}. Since IP address is used in estimating user geolocation when a CSAM seeking query is issued, this presents two edge cases:

\begin{itemize}
    \item \textbf{Treatment group traffic appearing to be from a convenience comparison country:} A user assigned to the Treatment cohort (e.g., United States) may utilize a VPN to route traffic through a comparison region (e.g., Norway). In this scenario, the user is analyzed as part of the convenience comparison user experience (legacy or no messaging) even when they are physically present in the treatment group country.
    
    \item \textbf{Convenience comparison group traffic appearing to be from a Treatment group country:} Conversely, a user from a Convenience comparison group country (e.g., Norway) may route traffic through a Treatment region (e.g., United States). These users are analyzed as part of the treatment group even if they are present physically in a convenience comparison country.
\end{itemize}

These scenarios are depicted graphically in Supplemental Material in Table A2. Both circumstances should theoretically lead to a depression of the computed effect sizes and thus our computed reduction in CSAM-seeking should be viewed as a conservative estimate. Users who employ VPNs to obfuscate their activity / geo-location also represent a segment of the population that is likely more technically sophisticated and highly motivated than the general population.
\subsection{Areas for Future Research}

This data has strong external validity, as it examines the actual behavior of Google Search users. Findings are, however, limited due to platform policies, data availability, and the way in which the new Onebox experience was launched. Additionally, while we could isolate the effect on Google Search, we were unable to measure behaviors off of Search. A user who terminated their session immediately after seeing the messaging looks identical in the logs data to a user who terminated their Google Search session and opened a Tor browser. Future studies researchers might explore feasible designs capable of measuring the sustained impact of interventions, or evaluating displacement to other platforms. 

Research with at-risk users might consider asking specific questions about how they would react to receiving the Onebox messaging. For example, is the messaging likely to stop them from seeking out CSAM at least in the immediate term? Would they interact with the help-seeking resources? If not, why not? 

Finally, ESPs implementing similar messaging or interventions might consider doing so in a way that allows for a more straightforward statistical analysis, such as conducting a randomized controlled experiment. This would simplify conclusions and strengthen claims. 

\section{Conclusion}
Directing more efforts towards early intervention serves a number of purposes. Most critically, primary and secondary prevention can reduce overall victimization and the serious downstream negative effects of victimization. Early intervention can also help ameliorate the current strain on the online sexual abuse ecosystem \citep{stanford_cybertipline_2024}, including NGOs such as NCMEC and IWF, and reduce the financial costs associated with child sexual abuse \citep{letourneau2018economic}. In our study, we found reorienting an existing CSAM intervention on Google Search to focus on help-seeking resulted in a significant decrease in on-seeking of CSAM and a significant increase in traffic to help-providing resources. We note that  the services were localized to the country of the user, often because the services themselves involve person-to-person assistance and/or language competency. An important implication of this work, particularly in light of recognition that CSAM is an exponentially growing problem, is that resources should be allocated to expanding coverage though translating and internationalizing services.

\section{Acknowledgments}
We would like to thank Amanda Storey, Zoe Darme, Sara Wiant, Charles Bradley, and Luke Coleman for their support in hosting the workshop. We would like to thank Professor Maggie Brennan for her valuable contributions during the initial phases of this project. We would like to thank Nikola Todorovic and Grigory Rozhdestvenskiy for their assistance in reviewing the manuscript. We would like to thank the helplines, particularly those who contributed data to support these analyses. 

\section{Funding disclosure}
Part of Caoilte Ó Ciardha's contribution to the study was funded by the Tech Coalition Safe Online Research Fund (Grant No. 23-EVAC-0015.2-University of Kent).
\section{Data availability}
Internal Google data cannot be made publicly available because they contain commercially sensitive information. Similarly, data from the helplines are withheld in accordance with their strict privacy policies.

\bibliographystyle{ACM-Reference-Format}
\bibliography{bibliography}
\newpage
\setcounter{table}{0} 
\renewcommand{\thetable}{A\arabic{table}} 
\begin{table}[!htbp]
\centering
\begin{tabular}{cc}
\hline
\textbf{Week (2025)} & \textbf{Number of visitors} \\
\hline
1  & 32  \\
2  & 67  \\
3  & 77  \\
4  & 63  \\
5  & 96  \\
6  & 83  \\
7  & 132 \\
8  & 163 \\
9  & 230 \\
10 & 149 \\
11 & 176 \\
12 & 220 \\
13 & 192 \\
14 & 215 \\
15 & 195 \\
16 & 169 \\
17 & 162 \\
18 & 125 \\
19 & 190 \\
20 & 144 \\ \hline
\end{tabular}
\caption{Dispositif STOP Web Traffic Data for 2025}
\label{tab:visitor_data}
\end{table}
\renewcommand{\thetable}{A\arabic{table}} 
\begin{table*}[!htbp]
  \caption{Impact of VPN-Mediated Traffic on Experimental Cohort Assignment and Effect Estimation. Because sampling is based on Query-Level Geolocation (IP Address), a user's physical location (Origin) may differ from their analyzed location (VPN Exit).}
  \label{tab:vpn_impact}
  \small
  \begin{tabularx}{\textwidth}{@{}
      >{\raggedright\arraybackslash}p{2.6cm} 
      >{\raggedright\arraybackslash}p{2.6cm} 
      >{\raggedright\arraybackslash}p{2.4cm} 
      >{\raggedright\arraybackslash}p{2.4cm} 
      >{\raggedright\arraybackslash}X 
  @{}}
    \toprule
    \textbf{Physical Origin} & \textbf{VPN Exit (IP)} & \textbf{Analysis Group} & \textbf{Intervention Received} & \textbf{Likely Impact on Effect Size} \\
    \midrule

    \textbf{Treated Region$_{A}$} \newline (e.g., USA) & 
    \textbf{Treated Region$_{B}$} \newline (e.g., UK) & 
    Treatment & 
    New Onebox$_{B}$ & 
    \textbf{Neutral / Underestimation.} User is analyzed as treated originating from treated Region B. However, users may not interact with the Onebox because it is localized for Region B (e.g., wrong phone numbers or translations), potentially depressing help-seeking rates despite successful deterrence. \\
    \addlinespace

    \textbf{Treated Region$_{A}$} \newline (e.g., USA) & 
    \textbf{Convenience Comparison Region} \newline (e.g., Norway) & 
    Convenience Comparison & 
    Legacy / None & 
    \textbf{Neutral / Dilution.} Target population is attributed to the Convenience Comparison group. They receive no intervention; their continued searching is expected and does not affect the Treatment stats. \\
    \addlinespace

    \textbf{Convenience Comparison Region} \newline (e.g., Norway) & 
    \textbf{Treated Region$_{A}$} \newline (e.g., USA) & 
    Treatment & 
    New Onebox$_{A}$ & 
    \textbf{Neutral / Underestimation.} A convenience comparison group user is attributed to the treatment dataset. If they receive the warning, the therapeutic resources offered are for the wrong region (e.g., US hotlines). If the user desists after encountering the new Onebox, they do not inflate the success rate. If this lack of relevant local help reduces the user's ability to desist, it inflates the treatment group's failure rate. \\
    \addlinespace

    \textbf{Convenience Comparison Region$_{A}$} \newline (e.g., Norway) & 
    \textbf{Convenience Comparison Region$_{B}$} \newline (e.g., Sweden) & 
    Convenience Comparison & 
    Legacy / None & 
    \textbf{Neutral.} User is correctly analyzed as convenience comparison. \\
    \bottomrule
  \end{tabularx}
\end{table*}
\end{document}